\pdfoutput=1

\documentclass[12pt,reqno]{article}
\usepackage{jheppub}
\usepackage{amsmath,amssymb,amsfonts}

\usepackage{graphicx}
\usepackage{epsfig}
\usepackage{dcolumn}
\usepackage{upgreek}
\usepackage{setspace}
\usepackage{enumitem}
\usepackage{array,multirow,bigdelim}

\title{Probing Scalar Effective Field Theories with the Soft Limits of Scattering Amplitudes}

\author[a]{Antonio Padilla,}
\author[b]{David Stefanyszyn}
\author[a]{and Toby Wilson}
\affiliation[a]{School of Physics and Astronomy, University of Nottingham, Nottingham NG7 2RD, UK}
\affiliation[b]{Van Swinderen Institute for Particle Physics and Gravity, University of Groningen, Nijenborgh 4, 9747 AG Groningen, The Netherlands}

\emailAdd{antonio.padilla@nottingham.ac.uk, d.stefanyszyn@rug.nl, toby.wilson@nottingham.ac.uk}

\abstract{We investigate the soft behaviour of  scalar effective field theories (EFTs) when there is a number  of  distinct derivative power counting parameters, $\rho_1< \rho_2<\ldots < \rho_Q$. We clarify the notion of an enhanced soft limit and use these to extend the scope of on-shell recursion techniques for scalar EFTs. As an example, we perform a detailed study of theories with two power counting parameters, $\rho_1=1$ and  $\rho_2=2$, that include the shift symmetric generalised galileons. We demonstrate that the minimally enhanced soft limit uniquely picks out the Dirac-Born-Infeld (DBI)  symmetry, including DBI galileons. For the exceptional soft limit we  uniquely pick out the special galileon  within the class of theories under investigation. We study the DBI galileon amplitudes more closely, verifying the validity of the recursion techniques in generating the six point amplitude, and explicitly demonstrating the invariance of all amplitudes under DBI galileon duality.}

\begin{document}
{\setstretch{1}
\maketitle
}
\section{Introduction} \label{intro}
Built upon key physical principles such as Lorentz invariance and unitarity, the modern S-matrix program has revealed remarkable new structures  within gauge theories and gravity, previously hidden from standard Lagrangian methods  (for a review, see \cite{Elvang}). In seemingly unrelated work, cosmological model builders have recently been exploiting a new class of scalar effective field theories (EFTs), dubbed {\it galileons} \cite{galileons}, in order to tackle a whole host of cosmological issues, from the nature of dark energy to the initial singularity (for a review, see \cite{review}).  An unlikely connection between these two  distinct research directions was made by Cachazo, He and Yuan, whose simple S-matrix constructions included the so-called {\it special galileon} as an example \cite{CHY}. When applied to late time cosmology, galileons often suffer from a dangerously low unitarity cut-off (see e.g. \cite{EFTforMG, Luty, Vain}), with question marks over one's ability to perform a consistent  UV completion \cite{Nima}.  However, their appearance in the CHY scattering equations and possible connections to ambitwistor constructions \cite{ambi} now offer some hope of understanding some of  these phenomenologically interesting field theories on a more fundamental level.

The status of the special galileon amongst scalar EFTs was further enhanced by a detailed study of  scattering amplitudes and their soft limits \cite{soft}. The special galileon is an example of an {\it exceptional} EFT whose soft limits are maximally enhanced \cite{soft2}. Dirac-Born-Infeld (DBI) theory\cite{DBI1,DBI2} is another example. Scalar EFTs with enhanced soft limits admit on-shell recursion relations that allow one to construct all higher order tree level amplitudes from a small number of lower point amplitudes \cite{recursion, soft2} (see \cite{luo} for further generalisations), reminiscent of the  BCFW recursion relations originally developed for Yang-Mills theory \cite{BCFW}.  The exceptional EFTs are the most economical in this regard, since their couplings only depend on a single parameter, the structure of the interactions is protected by an enhanced symmetry that can be directly related to the soft limit using Ward identities \cite{soft2}. For the case of the special galileon, this extra symmetry was identified in \cite{hidden}.

The soft limit of scattering amplitudes is realised by taking the momentum $p$ of one external leg soft i.e. expanding the amplitude around $p=0$. When applied to 
pion scattering amplitudes, one encounters the well known {\it Adler zero} \cite{adler1,adler2},  which states that the amplitude for emission of a single soft pion should vanish\footnote{See the first appendix of \cite{soft2} for a concise review of the Adler zero.}.  As explored in detail in \cite{soft,soft2}, we can consider generalisations of the Adler zero for which  the scattering amplitude scales as 
\begin{equation} \label{softamp}
\mathcal{A}(p) = \mathcal{O}(p^{\sigma})
\end{equation} 
in the soft limit, with $\sigma$  a positive integer.  The soft limit is regarded as {\it enhanced} if the value of $\sigma$ exceeds the naive value expected from dimensional analysis, and a simple counting of derivatives per field. This enhancement suggests  hidden structure in the scattering  amplitudes related to the existence of additional symmetries.

In \cite{soft,soft2} a classification scheme for scalar EFTs was developed, centred on the behaviour of their scattering amplitudes in the infrared. The theories under consideration were characterized by four parameters, $(\rho, \sigma, d, v)$, related to the number of derivatives per interaction, the soft behaviour of the amplitude, the spacetime dimension, and the valency of the leading interaction respectively. In particular, to enforce the cancellations between tree level Feynman diagrams required for 
an enhanced soft limit, a unique power counting condition was imposed on the interactions, corresponding to Lagrangians with the schematic form
 \begin{equation} \label{softlag}
\mathcal{L}_{(\rho)} = -\frac12 (\partial \phi)^{2} +(\partial \phi)^{2}\sum_{n=1}^{\infty}  c_{n} \partial^{m}\phi^{n}
\end{equation} 
where $c_{n}$ are coupling constants. By Lorentz invariance we require  the number of additional derivatives, $m=\rho n$ to be even, with the power counting parameter, $\rho$,  given by a fixed non-negative rational number. One can  impose enhanced soft limits in general tree level scattering amplitudes by enforcing constraints on the coupling constants of the underlying theory. These constraints can ultimately be used to identify new extended symmetries and new structures within scalar EFTs.

An enhanced soft limit is caused by cancellations between tree level Feynman diagrams of different topologies. For two diagrams to be able to cancel at least in principle, they must have an equivalent number of external legs, such that they contribute to the same scattering process, and by dimensional analysis have the same scaling with momenta. Consider one diagram with $A$ vertices where the $i^{\text{th}}$ vertex has $n_{i} + 2$ legs and $\rho_{i}n_{i} + 2$ powers of momenta and similarly another diagram with $B$ vertices where the $j^{\text{th}}$ vertex has $n_{j}+2$ legs and $\rho_{j}n_{j} +2$ powers of momenta. These tree level diagrams respectively have $A-1$ and $B-1$ internal lines and therefore matching the number of external legs is equivalent to
\begin{equation} \label{conditionlegs}
\sum_{i=1}^{A}n_{i} = \sum_{j=1}^{B}n_{j}
\end{equation}
and following the same logic for matching the powers of momenta yields 
\begin{equation} \label{conditionmomenta}
\sum_{i=1}^{A}\rho_{i}n_{i} = \sum_{j=1}^{B}\rho_{j}n_{j}.
\end{equation}
The most simple solution to this system is that every operator has a common value of $\rho$ such that all diagrams contributing to a given process share a common scaling with momenta allowing for \textit{any} two diagrams to in principle cancel. This class of theories are the ones studied in \cite{soft,soft2}. However, there are other solutions corresponding to theories which can indeed realise enhanced soft limits but where operators do not share a common $\rho$. For example, consider a 6-point amplitude whose structure is dictated by the following schematic Lagrangian
\begin{equation}
\mathcal{L} = -\frac{1}{2} (\partial \phi)^{2} + g_{1} (\partial \phi)^{4} + g_{2} (\partial \phi)^{2} \partial^{4} \phi^{2} + g_{3}(\partial \phi)^{4} \partial^{4} \phi^{2}.
\end{equation} 
Written in the form of (\ref{softlag}), each interaction has a different value of $\rho$, namely: $1$,$2$ and $6/4$ respectively. The contact diagram with a single $g_{3}$ vertex scales with the same power of momenta as the diagram built from a $g_{1}$ vertex connected by a single propagator to a $g_{2}$ vertex, so these two diagrams have a chance of producing some cancellations. 

These generalised cancellations open up the possibility of finding new  structures within a larger class of scalar EFTs. The main purpose of this paper is to investigate precisely these possibilities. Indeed, one of our original motivations for doing this was to ask the question: how special is the special galileon? Is it the unique theory with that particular soft behaviour, or is it part of a larger family? Extensions involving additional fields were proposed in \cite{extensions}. Within the class of double $\rho$ theories with $\rho_1=1, \rho_2=2$ describing a single scalar field,  we will show that the special galileon is indeed the unique theory with a $\sigma=3$ soft limit. Note that this class of theories includes the shift symmetric generalised galileons \cite{george}  (aka Horndeski theories).  As well as showing the uniqueness of the special galileon, we will demonstrate that the $\sigma=2$ soft limit uniquely corresponds to the so-called DBI galileon theories \cite{dbigal}, which share the same symmetries as plain old DBI.  We will study the DBI galileon amplitudes in greater detail,  verifying the validity of recursion techniques in generating higher point amplitudes from the seeds of lower point amplitudes, as well as explicitly demonstrating their  invariance under the recently discovered DBI galileon duality \cite{extremerel}.

The rest of this paper is organised as follows.  In section \ref{oldresults}, we will review some aspects of the single $\rho$ classification scheme developed in \cite{soft,soft2}. This will then be generalised to multiple $\rho$ theories in section \ref{newresults}, with particular emphasis on the double $\rho$ case. In section \ref{example}, we will explore soft limits of scattering amplitudes within the class of  double $\rho$  theories with $\rho_1=1, \rho_2=2$. As stated above, we will explicitly demonstrate the uniqueness of the DBI galileon and the special galileon as theories with soft limits enhanced up to $\sigma=2$ and $\sigma=3$ respectively. In section \ref{sec:dbigal}, we will study DBI galileon amplitudes in greater detail, checking the validity of recursion relations and demonstrating their invariance under DBI galileon duality. We summarize our results in section \ref{summary}.

\section{Single $\rho$ theories: DBI, galileons and the special galileon} \label{oldresults}
Here we will briefly review the main results for single $\rho$ theories presented in \cite{soft,soft2}. As our starting point we take the Lagrangian  (\ref{softlag}) focussing on the most interesting cases of $\rho=1$ and $\rho=2$ i.e. where $m=n$ and $m=2n$ respectively. For these single $\rho$ theories the Lagrangian $\mathcal{L}_{(\rho)}$ is reduced to $\mathcal{L}_{(\rho;\sigma)}$ which is the subset with soft amplitudes of the form (\ref{softamp})\footnote{Note that we adopt a slightly different notation to that of \cite{soft,soft2} in order to avoid confusion when we consider multiple $\rho$ theories.}.  We will also assume that we are in $3+1$ dimensions, unless we explicitly state otherwise.  

In each case an enhanced soft limit satisfies
\begin{equation} \label{sigmabound}
\sigma > \frac{\rho n+2}{n+2}
\end{equation} 
since the other regime is guaranteed counting the number of derivatives per fields. So for $\rho=1$, where the Lagrangian is
\begin{equation} \label{rhoisone}
\mathcal{L}_{(1;1)} = -\frac{1}{2} (\partial \phi)^{2} + \sum_{n=2}^{\infty} c_{n}(\partial \phi)^{2n},
\end{equation}
we are interested in soft limits with $\sigma>1$. We remind the reader that if the only symmetry is a constant shift of the scalar, the soft limit degree is $\sigma=1$ due to the Adler zero. Given that $\sigma$ is an integer, the first interesting case is where $\sigma =2$ and constraining $c_{n}$ such that the theory realises this enhanced soft limit yields the unique subset of (\ref{rhoisone}) corresponding to
\begin{equation} \label{DBI}
\mathcal{L}_{(1;2)} = -\lambda \sqrt{1 + \frac{(\partial \phi)^{2}}{\lambda}}  
\end{equation}
for positive or negative $\lambda$. This is the scalar sector of the DBI Lagrangian where the infinite number of coupling constants has been reduced to a single parameter $\lambda$. The symmetry related to the quadratic soft limit degree is
\begin{equation} \label{dbisymm}
\phi \rightarrow \phi + c + v_{\mu}x^{\mu} + \phi v^{\mu}\partial_{\mu}\phi
\end{equation}
where $v^{\mu}$ is a constant vector. This symmetry can be understood by considering a dynamical 3-brane embedded in a five dimensional Poincare invariant bulk where $\phi(x)$ is the co-ordinate in the extra dimension. With only a tension term on the brane the effective action is (\ref{DBI}). Full details of this probe brane construction can be found in many places in the literature e.g. \cite{dbigal,extremerel,embeddedbranes} so we do not repeat it here.
For the $\rho=1$ class of Lagrangians the story stops here since there is no way to realise an enhanced soft limit with $\sigma >2$. All higher order constraints simply reduce the S-matrix to the trivial limit of a free theory. The action (\ref{DBI}) therefore represents the maximal enhancement for a $\rho=1$ theory and as we discussed in the introduction, is an exceptional theory.

For $\rho=2$ the Lagrangian takes the schematic form
\begin{equation} \label{rhoistwo}
\mathcal{L}_{(2;1)} = -\frac{1}{2} (\partial \phi)^{2} + (\partial \phi)^{2}\sum_{n=1}^{\infty} (\partial^{2} \phi)^{n}
\end{equation}
where in contrast to the $\rho=1$ Lagrangian above where at each order in $n$ there was a single free coupling constant, there are many free coupling constants due to the different ways the derivatives can be contracted. In any case, again by equation (\ref{sigmabound}) the first enhancement occurs when $\sigma=2$ and demanding that the theory (\ref{rhoistwo}) realises a quadratic soft limit degree induces constraints on the coupling constants which ultimately yields the galileon Lagrangian whose form in four spacetime dimensions is \cite{galileons}
\begin{equation} \label{gallag}
\mathcal{L}_{(2;2)} = -\frac{1}{2} (\partial \phi)^{2} + (\partial \phi)^{2} \sum_{n=1}^{3} c_{n} \text{det}_n
\end{equation}
where $ \text{det}_n=n! \partial^{[\mu_1}\partial _{\mu_1} \phi \ldots \partial^{\mu_n]}\partial _{\mu_n}\phi$.
The enhanced soft limit is related to the galileon global symmetry $\phi \rightarrow \phi + b_{\mu}x^{\mu} + c$ which was first seen in the decoupling limit of DGP gravity \cite{DGP}. Even though the cubic galileon term is indeed invariant under the galileon symmetry, one can consistently set its co-efficient to zero without changing the theory thanks to the galileon duality \cite{unification,duality2}. In this sense the infinite number of coupling constants of (\ref{rhoistwo}) have been reduced to a pair of galileon coupling constants. 

We can further constrain the space of coupling constants by requiring that the amplitude has $\sigma=3$ behaviour in the soft limit to yield a theory $\mathcal{L}_{(2;3)}$ which is equivalent to (\ref{gallag}) but with $c_{1} = c_{3} = 0$ i.e. the quartic or {\it special} galileon limit where all amplitudes with an odd number of particles vanish. The symmetry associated with this further enhancement was identified in \cite{hidden} and is an extension of the galileon symmetry with quadratic dependence on the co-ordinates and the first derivative of the scalar. Specifically it is
\begin{equation} \label{specgalsymm}
\phi \rightarrow \phi + s_{\mu\nu}x^{\mu}x^{\nu} + 12 c_{2} s^{\mu\nu}\partial_{\mu}\phi \partial_{\nu}\phi
\end{equation} 
where $s_{\mu\nu}$ is symmetric and traceless.  A generalisation of the Adler zero derivation can relate the field independent part of this symmetry directly to the soft limit \cite{soft2}. The story for $\rho=2$ ends at $\sigma=3$ since again any higher order constraints simply reduce the theory to the free limit. The special galileon is another exceptional field theory, representing maximal enhancement within the $\rho=2$ class.   

Let us conclude this section by stating the main results that arose from the systematic study of  single $\rho$ theories \cite{soft2}
\begin{itemize}
\item the soft limit degree is bounded by the number of derivatives per interaction, specifically, $\sigma \leq \rho+1$. This bound is saturated by the exceptional EFTs.
\item the soft limit degree of every non-trivial EFT is strictly bounded by $\sigma \leq 3$. Arbitrarily enhanced soft limits are forbidden.
\item Non-trivial soft limits require that the valency of the leading interaction is bounded by the spacetime dimension, specifically, $v \leq d+1$. Note that for derivatively coupled theories, kinematic considerations forbid $v=3$, so we require $v \geq 4$.
\end{itemize}
\section{Beyond single $\rho$ theories} \label{newresults}
Our goal is to investigate the soft behaviour of scalar EFTs with more than one derivative power counting parameter. To this end we consider a generic multi $\rho$ theory described by a Lagrangian with the following schematic form
\begin{equation} \label{multirho}
\mathcal{L}_{(\rho_{1},\ldots,\rho_{Q})} =-\frac12  (\partial \phi)^{2} + (\partial \phi)^{2} \sum_{n_1+\ldots +n_Q\geq 1} c_{n_{1}, \ldots, n_{Q}} (\partial^{\rho_{1}} \phi)^{n_{1}} \ldots (\partial^{\rho_{Q}} \phi)^{n_{Q}}
\end{equation}
where $Q$ describes the number of distinct $\rho$, and we assume without loss of generality that $\rho_1<\rho_2<\ldots <\rho_Q$. We shall refer to a theory of this type as a $(\rho_{1},...,\rho_{Q})$ theory.  Lorentz invariance tells us that $\sum_{k=1}^{Q} \rho_{k}n_{k}$ is even. In comparison to the single $\rho$ case studied in \cite{soft}, our aim here is to classify theories of the form (\ref{multirho}) based on their behaviour in the soft limit yielding the subsets $\mathcal{L}_{(\rho_{1},\ldots,\rho_{Q};\sigma)}$.  Based on the total number of derivatives per field, an enhanced soft limit is one for which 
\begin{equation}
\sigma > \frac{2 + \sum_{k=1}^Q \rho_{k}n_{k} }{2 + \sum_{k=1}^Q n_{k} } \qquad \forall n_k.
\end{equation}
For an $N$ point amplitude, the right hand side of this inequality is maximised when all fields see the largest $\rho$, ie $n_k=0$ for $k=1, \ldots Q-1$ and $n_Q=N-2$. Then the condition for an enhanced soft limit becomes
\begin{equation} \label{softcond}
\sigma >\rho_Q+ \frac{2}{N}(1-\rho_Q) \qquad \forall N \geq 4
\end{equation}
and assuming $\rho_Q>1$, it follows that a non-trivial soft limit corresponds to 
\begin{equation}
\sigma \geq \rho_Q> \rho_{Q-1}> \ldots > \rho_1.
\end{equation}

To achieve an enhanced soft limit, we need cancellations between Feynman diagrams. As explained in the introduction, this can happen even in theories without a common $\rho$. However, the primary difference, compared with single $\rho$ theories, is that cancellations cannot take place between all pairs of diagrams, instead the diagrams contributing to a given process will naturally split into groups depending on how they scale with momenta with cancellations only possible within individual groups.  Now to enforce cancellations between two diagrams of different topologies we must again match the number of external legs and the overall scaling with momenta for each diagram. Generalising the conditions for cancellation we outlined in section \ref{intro}, we let the first diagram have $A$ vertices where the $i^{\text{th}}$ vertex has $\sum_{k=1}^{Q}n_{i,k} +2$ legs, where $n_{i,k}$ is the number of legs contributing to the $i^{\text{th}}$ vertex carrying $\rho_{k}n_{k}$ derivatives, and $\sum_{k=1}^{Q} \rho_{k}n_{i,k} +2$ powers of momenta and similarly the second diagram have $B$ vertices where the  $j^{\text{th}}$ vertex has $\sum_{k=1}^{Q}n_{j,k} +2$ legs and $\sum_{k=1}^{Q} \rho_{k}n_{j,k} +2$ powers of momenta. Now matching the number of external legs yields 
\begin{equation} \label{conditions1}
\sum_{i=1}^{A}  \sum_{k=1}^{Q} n_{i,k} = \sum_{j=1}^{B} \sum_{k=1}^{Q}  n_{j,k} 
\end{equation}
while matching the scalings with momenta is equivalent to
\begin{equation}  \label{conditions2}
\sum_{k=1}^{Q} \left(\rho_{k} \sum_{i=1}^{A} n_{i,k}\right) = \sum_{k=1}^{Q} \left( \rho_{k} \sum_{j=1}^{B} n_{j,k} \right).
\end{equation}
For the case of $Q=1$, these conditions reduce to (\ref{conditionlegs},\ref{conditionmomenta}) as expected and correspond to the theories studied in \cite{soft}. Of interest in this paper are theories with $Q  \geq 2$. 

An important ingredient in the systematic study of single $\rho$ theories were the on-shell recursion relations described in \cite{recursion,soft2}. These can be adapted to the multi $\rho$ case with a minimum of fuss.  To this end, consider an $N$ point amplitude, ${\cal A}_N(p_1, p_2, \ldots, p_N)$, and perform an {\it all but r} rescaling of the external momenta $p_i \to p_i(z)$ where 
\begin{equation}
 p_i (z)= \begin{cases} p_i (1-z\alpha_i) & \textrm{ for  $i=1, .. , N_s $} \\ p_i+z q_i  &  \textrm{ for  $i=N_s+1, .. , N$}\end{cases}
\end{equation}
and $N_s=N-r$.  The transformation parameters $(\alpha, q)$ are constrained by  momentum conservation 
\begin{equation}
\sum_{i=1}^{N_s} \alpha_i p_i=\sum_{i=N_s+1}^N q_i,
\end{equation}
and on-shell conditions
\begin{equation}
 q_i^2=q_i \cdot p_i=0 \qquad i=N_s+1, \ldots N.
 \end{equation}
 In $d$ spacetime dimensions, the $(\alpha, q)$ correspond to a set of $N_s+rd$ free parameters, with a total of $d+2r$ constraints.  There are also the following redundancies in our parametrization 
 \begin{equation}
 \delta \alpha_i=\lambda \alpha_i, ~\delta q_i=\lambda q_i; \qquad \delta \alpha_i=\mu, ~\delta q_i=-\mu p_i 
 \end{equation}
where $\lambda$ and $\mu$ are arbitrary constants. This reduces the number of true degrees of freedom by two, so in order to not overconstrain the system of rescalings, we require
\begin{equation}
N_s+rd-2 \geq d+2r.
\end{equation}
It follows that these rescalings can only probe those amplitudes for which
\begin{equation} \label{condN}
N \geq d+2-r(d-3)
\end{equation}
and, of course, 
\begin{equation}
N \geq r.
\end{equation}
As usual the rescalings endow the amplitude with dependence on the complex deformation parameter $z$, ${\cal A}_N \to {\cal A}_N(z)$. To find the recursion relation, we consider 
\begin{equation} \label{intA}
\oint_C\frac{A_N(z)}{z F_N(z)} dz=0
\end{equation}
where $C$ is a large contour at infinity and we define
\begin{equation}
F_N(z)=\prod_{i=1}^{N_s} (1-z\alpha_i)^\sigma
\end{equation}
where $\sigma$ is the soft degree of the relevant EFT. The integral is assumed to vanish thanks to the asymptotic behaviour of the integrand at large $z$. To see the condition for this to happen, we note that $F_{N}(z) \sim  z^{\sigma N_s}$ and ${\cal A}_N(z) \sim z^{m+2}$, at large $z$, where a simple counting of derivatives yields, $m=\sum_{k=1}^Q \rho_k n_k$.  To avoid the undesired pole in the integrand at infinity, we require $m+2<N_s \sigma$, for all possible combinations of $n_k$. Note that once again the left hand side of this inequality is maximised when all fields see the largest $\rho$.  It follows that a necessary and sufficient condition for rapid enough fall off at infinity is given by
\begin{equation} \label{condNs}
2+\rho_Q (N-2)<N_s \sigma.
\end{equation} 
For $r=0$, this yields an identical condition as the one for an enhanced soft limit (\ref{softcond}), which means a suitable rescaling can always be found for the theories of interest.  For larger values of $r$ the condition is stronger so these rescalings can only be used when there is an even greater level of enhancement in the soft limit.

When these conditions hold we can apply Cauchy's theorem to (\ref{intA}) to derive \cite{recursion, soft2}
\begin{equation} \label{recurs}
{\cal A}_N(0)=-\sum_{I} \textrm{Res}_{z_I^\pm} \left(\frac{{\cal A}_L(z) {\cal A}_R(z)}{zP^2_I(z)F_N(z)}\right)
\end{equation}
where $I$ denotes a factorization channel, and  $P_I(z)=\sum_{_i \in I} p_i(z)$ the rescaled momentum along that channel, which are, of course, linear in $z$.  $z_I^\pm$ are the roots of $P^2_I(z)=0$. Here we have used the fact that the would-be poles from $F_N(z)$ are cancelled by the zeros of the amplitude exactly on account of the soft limit. We have also used the fact that the other poles arise from factorization channels, where unitarity requires the residue to split into a product of lower point amplitudes ${\cal A}_L(z)$ and ${\cal A}_R(z)$. 

We could continue a detailed analysis as in \cite{soft2}, although it should now be clear that that all the results from \cite{soft2} will generalise to the multi $\rho$ case, provided we identify $\rho$ from \cite{soft2} with $\rho_Q$,  the largest power counting parameter when there are many different values in the set. Let us simply state the results, directly mirroring the single $\rho$ case:
\begin{itemize}
\item the soft limit degree is bounded by the number of derivatives per interaction, specifically, $\sigma \leq \rho_Q+1$, where $\rho_Q$ is the largest power counting parameter in the set.
\item the soft limit degree of every non-trivial EFT is strictly bounded by $\sigma \leq 3$. Arbitrarily enhanced soft limits are forbidden.
\item Non-trivial soft limits require that the valency of the leading interaction is bounded by the spacetime dimension, specifically, $v \leq d+1$. Note that for derivatively coupled theories, kinematic considerations forbid $v=3$, so we require $v \geq 4$.
\end{itemize}
\subsection{Double $\rho$ theories}
Let us now study some features of the $Q=2$ case a little more closely.  In particular the conditions (\ref{conditions1}) and (\ref{conditions2}) for cancellations between Feynman diagrams now reduce to 
\begin{eqnarray}
\sum_{i=1}^{A} (n_{i,1} + n_{i,2}) &=& \sum_{j=1}^{B} (n_{j,1} + n_{j,2}) \\
\rho_{1}\sum_{i=1}^{A} n_{i,1} + \rho_{2}\sum_{i=1}^{A} n_{i,2} &=& \rho_{1}\sum_{j=1}^{B} n_{j,1} + \rho_{2}\sum_{j=1}^{B} n_{j,2}. 
\end{eqnarray}
By construction we have $\rho_{1}  < \rho_{2}$, so the only solution to this system is
\begin{eqnarray}
\sum_{i=1}^{A}n_{i,1} &=& \sum_{j=1}^{B}n_{j,1} \\
\sum_{i=1}^{A}n_{i,2} &=& \sum_{j=1}^{B}n_{j,2}.
\end{eqnarray}
This simplification is not possible for $Q \geq 3$.

Let us construct the general form of the amplitudes for $(\rho_{1},\rho_{2})$ theories, based purely on kinematics and locality considerations. To do so we sum over the number of propagators. To introduce notation initially consider a contact diagram contribution to a $N$ point amplitude which given that $N = n_{1}+n_{2}+2$, has $\rho_{1}n_{1} + \rho_{2}(N-n_{1}-2) + 2$ powers of momenta. The amplitude is therefore constructed from $(\rho_{1}n_{1} + \rho_{2}(N-n_{1}-2))/2 + 1$ powers of $s_{ij}$ where
\begin{equation}
s_{ij} = (p_{i} + p_{j})^{2} = 2 p_{i} \cdot p_{j}
\end{equation}
are kinematic invariants making the Lorentz invariance of the S-matrix manifest. Now if we let $\alpha$ label pairs of external legs, the amplitude is of the form
\begin{equation}
\mathcal{A}_{N} = \sum_{n_{1} = 0}^{N-2} \sum_{\alpha} c_{\alpha}^{(0, n_{1})} \left(s_{\alpha_{1}} \ldots s_{\alpha_{X(n_{1})}}\right)
\end{equation}
where 
\begin{equation}
X(n_{1}) = \frac{\rho_{1}n_{1} + \rho_{2}(N-n_{1}-2)}{2} + 1
\end{equation}
and we are assuming that $\rho_{1}$ and $\rho_{2}$ are specified as required to define a $(\rho_{1},\rho_{2})$ theory. The generalisation of this for diagrams with $y$ propagators yields
\begin{equation}
\mathcal{A}_{N} = \sum_{n_{1} = 0}^{N-2} \sum_{\alpha,\beta} c_{\alpha}^{(y, n_{1})} \frac{\left(s_{\alpha_{1}} \ldots s_{\alpha_{X(n_{1})+y}}\right)}{s_{\beta_{1}}\ldots s_{\beta_{y}}}
\end{equation}
where $\beta$ labels factorisation channels whose propagators are fixed by energy conservation at each vertex. To construct the full general amplitude we sum over contributions from all possible number of propagators yielding 
\begin{equation} \label{AN}
\mathcal{A}_{N} = \sum_{y=0}^{N-3} \sum_{n_{1} = 0}^{N-2} \sum_{\alpha,\beta} c_{\alpha}^{(y, n_{1})} \frac{\left(s_{\alpha_{1}} \ldots s_{\alpha_{X(n_{1})+y}}\right)}{s_{\beta_{1}}\ldots s_{\beta_{y}}}.
\end{equation}
Finally, we note that unitarity considerations imply that the coefficients $c_\alpha^{(y, n_{1})}$ for $y \geq 1$ are related to coefficients appearing in lower point amplitudes. This follows on account of the fact that near a pole in the propagator the amplitude factorises into a pair of lower point amplitudes. 

\subsection{Computing soft limits in practice} \label{practice}
Let us briefly explain how to compute the soft degree of scattering amplitudes and impose enhancements when armed with a set of formulae of the form of (\ref{AN}). In all cases we are interested in, the first non-trivial amplitude is at 4-point since in a derivatively coupled theory of a single scalar the on-shell 3-point amplitude always vanishes. For a single scalar theory the 4-point amplitude is never too complicated and when expressed in terms of the usual Mandelstam variables it is simple to read off the soft degree and the conditions required for enhancement after sending one of the external momenta soft. Without loss of generality for any $N$-point amplitude with $p_{1} \ldots p_{N}$ external momenta we choose to send $p_{1}$ soft. This is a generic choice since the single scalar amplitudes we will consider are invariant under any exchanges of momenta for scattering of identical particles. For larger point processes the amplitudes quickly become complicated and the computations yield long expressions making computing the soft degree analytically a difficult task. This is compounded by the fact that in four spacetime dimensions there are only four linearly independent momentum vectors leading to Gram-determinant relations that fix the momenta with $N>4$ in terms of the first four. For 5-point processes the required constraint is simply momentum conservation but for 6-point processes and beyond extra constraints must be imposed to ensure that the soft limit is taken consistently. For these reasons it is more efficient to compute the amplitudes numerically using a redundant basis for the momenta. We define the first four external momenta in such a way that they are all on-shell and linearly independent of each other, for example, $p_1 = (1,0,0,1)$, $p_2=(1,0,1,0)$, $p_3=(1,1,0,0)$ and $p_4=(\sqrt{2},1,1,0)$, and fix the remaining external momenta as linear combinations of the first four momenta as $p_j = \sum_{i=1}^{4}\alpha_{ij}p_i$ for $j>4$. Some of the $\alpha_{ij}$ are fixed such that all external momenta are on-shell and to ensure that the system as a whole conserves energy and momentum. The remaining $a_{ij}$ are set randomly and one should check that any enhanced soft limits hold for a range of choices. To take the soft limit one should  send $p_{1} \rightarrow w p_{1}$ and expand the resulting numerical amplitudes for small $w$.

For all processes the resulting amplitudes are functions of the constants $c_\alpha^{(y, n_{1})}$, themselves  functions of the coupling constants whenever they are derived directly from a Lagrangian description. There are also  numerical factors coming from our choice of basis for the momenta, and the parameter $w$ which measures the degree of the soft limit. As explained above, for shift symmetric theories  all amplitudes will be at least linear in $w$ due to the Adler's zero condition. To realise enhanced soft limits of the form $w^{\sigma}$ one simply sets the coefficients of all lower order powers of $w$ to zero thereby placing constraints on the coupling constants. One should  repeat this process for different choices of momenta to yield constraints on the coupling constants which are momentum independent.

\section{An example: enhanced soft limits for $(1,2)$ theories} \label{example}
As an example of a non-trivial multi $\rho$ theory, let us focus on the $(1,2)$ class, with $\rho_1=1$ and $\rho_2=2$.  To probe the soft limits we can apply the methods described in the previous section to constrain the unknown coefficients $c_\alpha^{(y, n_{1})}$ in the amplitudes. This is non-trivial on account of the unitarity relations and disguises the precise relationship to known Lagrangians. A simple way to bypass these difficulties  is to start from a Lagrangian formulation and  to derive the formulae for the general amplitudes, which  will of course take the form of (\ref{AN}), but will now  encode information about the Lagrangian directly, from locality and unitarity to its couplings. That is the approach we will take here.

In general one would like to consider all possible vertices within a $(1,2)$ theory which would involve building operators out of $L_{\mu\nu} = \partial_{\mu}\phi \partial_{\nu}\phi$, $H_{\mu\nu} = \partial_{\mu}\partial_{\nu}\phi$ and traces, all containing at least one factor of $L_{\mu\nu}$\footnote{Interactions built purely from $H_{\mu\nu}$ cannot fall into the $(1,2)$ class.}. Up to five point interactions, these are given by a pure $\rho_1=1$ interaction
\begin{equation}
\text{Tr}[L]^{2} = (\partial \phi)^{4},
\end{equation}
a number of pure $\rho_2=2$ interactions
\begin{eqnarray}
\text{Tr}[L]\text{Tr}[H] &=& (\partial \phi)^{2} \Box \phi \\
\text{Tr}[L] \text{Tr}[H]^{2} &=& (\partial \phi)^{2} (\Box \phi)^{2}\\
\text{Tr}[L] \text{Tr}[H^{2}] &=& (\partial \phi)^{2} (\partial_{\mu}\partial_{\nu}\phi)^{2}\\
\text{Tr}[LH] \text{Tr}[H] &=& \partial_{\mu}\phi \partial_{\nu}\phi \partial^{\mu}\partial^{\nu}\phi \Box \phi\\
\text{Tr}[LHH] &=& \partial_{\mu}\phi \partial^{\mu} \partial_{\alpha}\phi \partial^{\alpha} \partial_{\nu}\phi\partial^{\nu}\phi \\
\text{Tr}[L] \text{Tr}[H]^{3} &=& (\partial \phi)^{2} (\Box \phi)^{3} \\
\text{Tr}[L] \text{Tr}[H] \text{Tr}[H^{2}] &=& (\partial \phi)^{2} \Box \phi (\partial_{\mu}\partial_{\nu}\phi)^{2} \\
\text{Tr}[L] \text{Tr}[H^{3}] &=& (\partial \phi)^{2} (\partial_{\mu}\partial_{\nu}\phi)^{3} \\
\text{Tr}[LH] \text{Tr}[H]^{2} &=& \partial_{\mu}\phi \partial_{\nu}\phi \partial^{\mu}\partial^{\nu}\phi (\Box \phi)^{2} \\
\text{Tr}[LH] \text{Tr}[H^{2}] &=& \partial_{\mu}\phi \partial_{\nu}\phi \partial^{\mu}\partial^{\nu}\phi (\partial_{\alpha}\partial_{\beta}\phi)^{2} \\
\text{Tr}[LHH]\text{Tr}[H] &=& \partial_{\mu}\phi   \partial_{\nu}\phi \partial^{\mu} \partial^{\alpha}\phi \partial_{\alpha} \partial^{\nu}\phi \Box \phi \\
\text{Tr}[LHHH] &=& \partial_{\mu}\phi  \partial_{\nu}\phi \partial^{\mu}\partial^{\alpha}\phi \partial_{\alpha}\partial_{\beta}\phi \partial^{\beta}\partial^{\nu}\phi , 
\end{eqnarray}
and finally a pair of  mixed interactions including both $\rho_1=1$ and $\rho_2=2$,
\begin{eqnarray}
\text{Tr}[L]^{2} \text{Tr}[H] &=& (\partial \phi)^{4} \Box \phi \\
\text{Tr}[L] \text{Tr}[LH] &=& (\partial \phi)^{2} \partial_\mu \phi \partial_\nu \phi \partial^\mu \partial^\nu \phi 
\end{eqnarray}
which are actually equivalent under integration by parts.

For this class of theories an enhanced soft limit corresponds to $\sigma \geq 2$, as shown in the previous section.  Working, for the moment, up to five point amplitudes, we note that the mixed interaction plays no role since it is a five point contact term that vanishes on-shell. This means we only have to worry about the pure $\rho_1=1$ interaction and the pure $\rho_2=2$ interactions. We now make the following observation: for any $N$-point amplitude, the group of diagrams with the largest scaling with momenta will always be those constructed from pure $\rho_2=2$ operators, since they  are the operators with the largest number of derivatives per field. Similarly, in the case where $N$ is even the group of diagrams with the smallest scaling with momenta will be constructed from purely $\rho_1=1$ operators since these have the fewest number of derivatives per field. Working up to five point amplitudes, this means our pure $\rho_1=1$ operators and our pure $\rho_2$ operators never work together to provide the cancellations required for enhanced soft limits. Since they must work independently we can immediately use the results of \cite{soft} to infer the following statements:
\begin{itemize}
\item for enhancement up to $\sigma=2$, the amplitudes up to five point must be a combination of the DBI amplitudes and the galileon amplitudes.
\item for enhancement up to $\sigma=3$, the amplitudes up to five point must be the special galileon amplitudes.
\end{itemize}
Once the four and five point amplitudes are given we can uniquely derive all of the higher point amplitudes using recursion relations.  This follows in $d=4$ dimensions from (\ref{condN}), which states that recursion relations can be used for $N \geq 6-r$. For $\sigma=2$ we see from (\ref{condNs}) that we must take $r=0$, whereas for $\sigma=3$ we take $r=1$. All that remains to be done is to identify the full theory, or more precisely the symmetry,  consistent with those higher amplitudes.  By studying up to seven point amplitudes within the class of shift symmetric generalised galileons, we will demonstrate that 
\begin{itemize}
\item for enhancement up to $\sigma=2$, the $(1,2)$ theory must exhibit DBI symmetry (\ref{dbisymm}). Within the class of shift symmetric generalised galileons this corresponds uniquely to the DBI galileon theory \cite{dbigal}.
\item for enhancement up to $\sigma=3$, the $(1,2)$ theory must exhibit the special galileon symmetry (\ref{specgalsymm}). Within the class of shift symmetric generalised galileons this corresponds uniquely to the special galileon.
\end{itemize}
To this end, we begin by writing the shift symmetric generalised galileons \cite{george}   in the following form 
\begin{equation} \label{gengal}
\mathcal{L}_{(1,2)} =-\frac12 (\partial \phi)^2+  \sum_{n \geq 1} a_n  (\partial \phi)^{2n+2}+\sum_{n\geq 1} \left( b_n  \textrm{det}_1+c_n   \textrm{det}_2 +d_n \textrm{det}_3 \right)(\partial \phi)^{2n} 
\end{equation}
where, with $H_{\mu\nu} = \partial_{\mu}\partial_{\nu}\phi$, we have
\begin{eqnarray}
\textrm{det}_1 &=& \text{Tr}[H] \label{det1} \\
\textrm{det}_2 &=& \text{Tr}[H]^{2} - \text{Tr}[H^{2}]\\
\textrm{det}_3 &=& \text{Tr}[H]^{3} + 2 \text{Tr}[H^{3}] - 3 \text{Tr}[H] \text{Tr}[H^{2}] \label{det3}
\end{eqnarray}
and $a_n, ~b_n, ~c_n$ and $d_n$ are (dimensionful)  constants to be constrained by the soft limits.  We can identify a number of interesting and known theories  within this class. For example, the K-essence, or $P(X)$ theories \cite{kessence} have $b_n=c_n=d_n=0$; galileons \cite{galileons} have $a_n=0, ~\forall n$ and $b_n=c_n=d_n=0, ~ \forall n\geq 2$;  and finally, the DBI galileons \cite{dbigal} have (see equation (\ref{dbigaleq}))
\begin{eqnarray}
a_n &=&-\frac12 \frac{(2n)!}{n!(n+1)!}   \left(-\frac{1}{4\lambda_0} \right)^n \nonumber \\
 b_n &=& \frac{\lambda_1}{2n} \left(-\frac{1}{\lambda_0} \right)^n\nonumber \\
   c_n &=&  \lambda_2 \frac{(2n)!}{(n!)^2}  \left(-\frac{1}{4\lambda_0} \right)^n \nonumber \\
    d_n&=& \lambda_3\left(-\frac{1}{\lambda_0} \right)^n. \nonumber
\end{eqnarray}
Our numerical analysis of the amplitudes reveals that up to seven point we have 
\begin{eqnarray}
{\cal A}_4 &=& a_1  \#_{4,4}  w^2+{\cal O}(w^3)  \\
{\cal A}_5 &=& (4 b_1^3+6 b_1 c_1+3 d_1) \#_{5,8}  w^2+{\cal O}(w^3) \\
{\cal A}_6 &=& \left[ (a_2 +4a_1^2)  \#_{6,6}  +  (8a_1 b_1^2+2b_1 b_2+6a_1 c_1+c_2) \#_{6,8}\right] w  \nonumber \\ &&  \qquad + \left[ a_1^2  \#'_{6,6}  +  (2b_1 b_2+c_2) \#'_{6,8}+a_1 b_1^2  \#''_{6,8} +a_1 c_1  \#'''_{6,8} \right] w^2+{\cal O}(w^3) \\
{\cal A}_7 &=& (2b_2c_1 +2b_1 c_2+d_2+8a_1 d_1+20a_1 b_1c_1+16 a_1 b_1^3+4 b_1^2 b_2)  \#_{7,10}  w  \nonumber \\ && \quad + \left[ a_1^2 b_1  \#_{7,8}  +  (2c_1 b_2+2b_1c_2+4b_1^2 b_2+d_2) \#'_{7,10}+a_1b_1^3   \#''_{7,10} +a_1 b_1 c_1  \#'''_{7,10}
\right. \nonumber  \\
&& \quad  \left. 
 +a_1 d_1 \#^{(4)}_{7,10}
+(6 b_1 c_1^2 +3b_1^2 d_1+3c_1 d_1+10 b_1^3 c_1+4 b_1^5)\#_{7, 12}
 \right] w^2 +{\cal O}(w^3) \hspace{1.05cm}
\end{eqnarray}
where we recall that $w$ is the soft parameter, as described in section \ref{practice}, and $\#_{n, m}, ~\#_{n, m}'$ etc denote distinct kinematic combinations of momentum entering the $n$ point amplitude and scaling with $m$ powers of momentum.  

To force an enhanced soft limit with $\sigma=2$ we set all the coefficients of all the distinct kinematic combinations to vanish, at order $w$.  This yields
\begin{eqnarray}
a_2 &=& -4a_1^2 \\
c_2 &=& -(8a_1 b_1^2+2b_1 b_2+6a_1 c_1) \\
d_2 &=& -(2b_2c_1 +8a_1 d_1+8a_1 b_1c_1).
\end{eqnarray}
Working up to seven point interactions, we can transform $b_2 \to -4 a_1 b_1$ and $b_3 \to \frac{64}{3} a_1^2 b_1$ by means of the following field redefinition
\begin{equation} \label{redef}
\phi \to \phi-(b_2+4 a_1 b_1) (\partial \phi)^4-\left(b_3+\frac{8}{3}a_1(b_2+4 a_1 b_1)\right) (\partial \phi)^6
\end{equation}
bringing the Lagrangian into standard form for the DBI galileon (see equation \ref{dbigaleq}). We will discuss this theory and its amplitudes in more detail in the next section. This now proves that the DBI symmetry uniquely gives rise to the $\sigma=2$ enhanced soft limit, within the class of theories with $\rho_1=1, ~\rho_2=2$.

To force a $\sigma=3$ soft limit, we must also set the relevant coefficients at order $w^2$ to vanish. This gives
\begin{equation}
a_1=a_2=0, \qquad c_2=-2b_1 b_2, \qquad d_2=-2c_1 b_2
\end{equation}
\begin{equation}
 4 b_1^3+6 b_1 c_1+3 d_1=0.
\end{equation}
The field redefinition (\ref{redef}), with $a_1=0$, can now be used to eliminate the $b_2$ and $b_3$ dependence, leaving us with a pure galileon Lagrangian characterised by 
\begin{equation} \label{class2}
d_1=-\frac23 b_1(2b_1^2+3c_1).
\end{equation}
Galileon Lagrangians are known to admit a duality transformation that allows us to identify seemingly different galileon theories \cite{unification,duality2}.  If we impose that the leading order kinetic term takes the  canonical form, galileon duality transformations will map the galileon interactions as $(b_1, c_1, d_1) \to (b_1(q), c_1(q), d_1(q))$ where
\begin{equation}
b_1(q)=b_1-\frac{q}{2}, \qquad c_1(q)=c_1-b_1 q-\frac{q^2}{4}, \qquad d_1(q)=d_1+c_1 q+\frac12 b_1 q^2-\frac{q^3}{12}. 
\end{equation}
Making use of these identifications, we recognise the class of Lagrangians (\ref{class2}) as special galileons ($c_1 \neq 0$ only). This now proves that the special galileon symmetry uniquely gives rise to the $\sigma=3$ enhanced soft limit within the  $\rho_1=1, ~\rho_2=2$ class. The freedom to perform galileon duality transformations without altering the amplitude serves as a good consistency check of our numerical results.

\section{DBI galileon amplitudes and recursion formulae}\label{sec:dbigal}
Let us now look at the amplitudes for the DBI galileon theory a little more closely. We begin by writing the DBI galileon Lagrangian in the following elegant form
\begin{equation} \label{dbigaleq}
{\cal L}_{DBI}=-\lambda_0 \gamma^{-1}+\lambda_1 \ln \gamma \textrm{det}_1+\lambda_2 \gamma \textrm{det}_2+\lambda_3 \gamma^2 \textrm{det}_3
\end{equation}
where 
\begin{equation}
\gamma=\frac{1}{\sqrt{1+\frac{(\partial \phi)^2}{\lambda_0}}}
\end{equation}
and $\textrm{det}_n$ are defined in equations (\ref{det1}) to (\ref{det3}). The four and five point scattering amplitudes are given by
\begin{eqnarray}
{\cal A}_4 &=&\frac{1}{4 \lambda_0} (s_{12}^2+s_{23}^2+s_{31}^2) +\frac{1}{2} \left(\frac{\lambda_1^2}{2 \lambda_0^2} -\frac{\lambda_2}{\lambda_0}\right)  (s_{12}^3+s_{23}^3+s_{31}^3)  \label{A4} \\
{\cal A}_5 &=& 24 \left( \frac{2\lambda_3}{\lambda_0}-\frac{\lambda_1 \lambda_2}{\lambda_0^2}+\frac{\lambda_1^3}{3\lambda_0^3}\right) \det(1,2,3,4)
\label{A5}
\end{eqnarray} 
where $\det(1, \ldots, n)=n!p_1^{[\mu_1} \ldots p_n^{\mu_n]} p_1{}_{\mu_1} \ldots p_n{}_{\mu_n}$. These are the seed amplitudes for all higher point amplitudes thanks to the recursion relations described in section \ref{newresults}.  For example, applying the recursion formulae (\ref{recurs}) with $\sigma=2$ and $r=0$ the six point point amplitude can be built from ten disinct factorization channels 
\begin{equation}
{\cal A}_6={\cal A}_6^{(123)}+\ldots
\end{equation}
where $\ldots$ denote the other factorization channels. Calculating the residues at the poles of the rescaled propagator, we have that \cite{recursion}
\begin{equation}
{\cal A}_6^{(123)}=\frac{1}{P_{123}^2}\left[\frac{{\cal A}_4^{(123)}(z^-) {\cal A}_4^{(456)}(z^-)}{F(z^-)\left(1-\frac{z^-}{z^+}\right)}+z^+ \leftrightarrow z^- \right]
\end{equation}
where $F(z)=\prod_{i=1}^6 (1-\alpha_i z)^2$, $z^\pm$ are the roots of $P_{123}^2(z)=0$, and 
\begin{multline}
{\cal A}^{(123)}_4(z)=\frac{1}{4 \lambda_0} (s^2_{12}(z)+s^2_{23}(z)+s^2_{31}(z)) \\
+\frac{1}{2} \left(\frac{\lambda_1^2}{2 \lambda_0^2}-\frac{\lambda_2}{\lambda_0} \right)  (s^3_{12}(z)+s^3_{23}(z)+s^3_{31} (z))  \qquad 
\end{multline}
where $s_{ij}(z)=(1-\alpha _iz) (1-\alpha_jz) s_{ij}$. We can obtain a similar formula for $ {\cal A}_4^{(456)}(z)$ by trading $123 \to 456$.  After some simplification we obtain
\begin{equation}
{\cal A}_6^{(123)}=\frac{B}{P_{123}^2} \sum_{\substack{i, j \in \{1,2, 3\} \\  k, l \in \{4, 5, 6\}}} T_{ijkl} (z^-) +z^+ \leftrightarrow z^- 
\end{equation}
where $B=\left(1-\frac{z^-}{z^+}\right)^{-1}$ and 
\begin{equation}
T_{ijkl}(z)=\frac{\left[\frac{s^2_{ij}}{4 \lambda_0}+\frac12 \left(\frac{\lambda_1^2}{2 \lambda_0^2} -\frac{\lambda_2}{\lambda_0}\right)  s^3_{ij} r_{ij}(z)\right] \Big[ ij \to kl \Big] }{\prod_{m \notin \{i,j,k,l\}} (1-a_m z)^2}.
\end{equation}
Here $r_{ij}(z)=(1-\alpha_i z)(1-\alpha_j z)$. One can happily check that the unphysical scaling parameters drop out of this expression and one is left with the correct formula for the six point amplitude.

We now conclude this section by verifying the invariance of our amplitudes under the DBI galileon duality discussed in \cite{extremerel}.   It is sufficient to prove the invariance of the four and five point amplitudes, the invariance of higher point amplitudes following directly from this and the recursion relations. Now, to adapt the results of \cite{extremerel}, we write our DBI galileon Lagrangian as
\begin{equation}
L_{DBI}=\sum_{i=1}^3  \frac{N_i}{\gamma} \det\left[ \delta^\mu_\nu+C_i \gamma \left(\partial^\mu\partial_\nu \phi-\gamma^2 \partial^\mu \phi (\partial^\alpha\partial_\nu \phi \partial_\alpha \phi\right)\right]
\end{equation}
where, after some integration by parts, we identify
\begin{equation}
\lambda_0=-\sum_i N_i, \qquad \lambda_1=\frac{-\sum_i N_i C_i}{\sqrt{\lambda_0}}, \qquad \lambda_2=\frac{-\sum_i N_i C_i^2}{2\lambda_0}, \qquad \lambda_3=\frac{-\sum_i N_i C_i^3}{12\lambda_0^{3/2}}.
\end{equation}
A duality transformation is a constant shift $C_i \to C_i+\nu$, or equivalently
\begin{eqnarray}
\lambda_0 &\to& \lambda_0 \\
\lambda_1 &\to &\lambda_1+\nu \sqrt{\lambda_0} \\
\lambda_2 & \to &  \lambda_2+\nu \frac{\lambda_1}{\sqrt{\lambda_0}}+\frac{\nu^2}{2} \\
\lambda_3 & \to & \lambda_3+\frac{\nu}{2}  \frac{\lambda_2}{\sqrt{\lambda_0}} +\frac{\nu^2}{4}\frac{\lambda_1}{\lambda_0}+\frac{\nu^3}{12}\frac{1}{\sqrt{\lambda_0}}.
\end{eqnarray}
We can now trivially verify that the amplitudes (\ref{A4}) and (\ref{A5}), and by recursion all higher point amplitudes,  are indeed invariant under this transformation.
\section{Summary} \label{summary}
We have initiated  a study of the soft behaviour of scalar effective field theories with multiple parameters required for derivative power counting. In particular, we have considered theories whose Lagrangians take the following schematic form
\begin{equation}
\mathcal{L}_{(\rho_{1},\ldots,\rho_{Q})} =-\frac12  (\partial \phi)^{2} + (\partial \phi)^{2} \sum_{n_1+\ldots +n_Q\geq 1} c_{n_{1}, \ldots, n_{Q}} (\partial^{\rho_{1}} \phi)^{n_{1}} \ldots (\partial^{\rho_{Q}} \phi)^{n_{Q}} \label{struc}
\end{equation}
where $Q$ describes the number  of distinct power counting parameters $\rho_1<\rho_2<\ldots <\rho_Q$.  Generically we find that the results and classifications of single $\rho$ theories \cite{soft2}, with $\rho=\rho_Q$,  also apply to multiple $\rho$ theories for which $\rho_Q$ is the largest power counting parameter. For example, a scalar EFT with a non-trivial soft limit is one whose soft degree is enhanced beyond the value naively expected from counting derivatives, $\sigma \geq \rho_Q$. This soft degree is also bounded from above, $\sigma \leq \rho_Q+1$ and $\sigma \leq 3$. Enhanced  soft limits happen thanks to cancellations between Feynman diagrams, although in contrast to the single $\rho$ case, for any given amplitude there are many contributions with different momentum scaling, requiring many distinct cancellations for each particular scaling. This enhanced soft behaviour also allows one to generalise the recursion formulae of \cite{recursion, soft2}, enabling us to generate all tree level amplitudes from a small number of lower point seed amplitudes.  Non-trivial soft limits also place  a bound on the valency of the leading interaction in terms of the spacetime dimension $v \leq d+1$, just as they do for single $\rho$ theories.

As an explicit example of multi $\rho$ theories we have considered the double $\rho$ case, with $\rho_1=1$ and $\rho_2=2$.  Note that this class of theories includes all shift symmetric generalised galileons (aka Horndeski theories), which is manifest when they are written in the elegant form given in \cite{george}.  By studying the behaviour of amplitudes in detail, we have been able to identify two non-trivial EFTs with enhanced soft behaviour.  For the minimal case, with soft degree $\sigma=2$,  we were able to uniquely identify the enhanced soft limit with the DBI symmetry, picking out the so-called DBI galileons \cite{dbigal} within the Horndeski class.  For the exceptional case with $\sigma=3$, we uniquely pick out the special galileon symmetry \cite{hidden}. This latter result satisfies one of the main motivations for our work in trying to establish ``how special is the special galileon?". 

We also study the amplitudes for DBI galileons in further detail. There are two seed amplitudes, at four and five point, which we compute explicitly, and demonstrate to be invariant under the DBI galileon duality discussed in \cite{extremerel}.  From these seed amplitudes one can use recursion relations to generate all higher point amplitudes, proving the duality invariance at tree level in all cases. We explicitly  calculate the six point amplitude using the recursion techniques and verify that they do indeed agree with the direct calculation using Feynman diagrams.

Finally, it is interesting to note that whilst this paper was in the final stages of preparation a detailed, but orthogonal study of the special galileon symmetry appeared \cite{Novotny}. There it was shown that the special galileon symmetry, or hidden galileon symmetry \cite{hidden}, is merely a fixed point of a new hidden galileon duality.  It stated that there are, in fact, a three parameter family of invariant galileon theories. This is only true when we admit a non-trivial tadpole -- if we forbid the tadpole by demanding a constant $\phi$ vacuum, the result is consistent with our claim regarding the uniqueness of the special galileon within the  $(\rho_1=1, \rho_2=2)$ class. The analysis of \cite{Novotny} also spells out a geometric construction for theories with the special galileon symmetry, analogous to the DBI brane constructions. This allows one to construct higher order invariant Lagrangians. To capture these we would need to extend our detailed analysis to higher $Q$, and possibly fractional values of $\rho$. This would be an interesting topic for future study in its own right.
\section*{Acknowledgments}
We would like to thank Tim Adamo, Jiri Novotny, Diederik Roest, Dimitri Skliros and Jaroslav Trnka for useful discussions. AP acknowledges the support of a Royal Society URF and STFC Joint Consolidated Grant, DS acknowledges the Dutch funding agency `Foundation for Fundamental Research on Matter' (FOM) for financial support and TW is funded by an STFC studentship.

\onehalfspacing

\end{document}